\documentclass[twocolumn,
preprintnumbers,amsmath,amssymb,prl]{revtex4-1}
\usepackage{amsmath,amssymb,,epsfig}%,amsthm}
\usepackage{dcolumn}% Align table columns on decimal point
\def\be{\begin{equation}}
\def\ee{\end{equation}}
\def\bea{\begin{eqnarray}}
\def\eea{\end{eqnarray}}
\def\d{\partial}
\def\L{\Lambda}
\def\half{\frac{1}{2}}
\def\a{\alpha}
\def\b{\beta}

\def\e{\epsilon}
\newcommand{\Tr}{\mbox{Tr}}

\begin{document}

\preprint{YITP-SB-10-39}

\title{Holographic dual of free field theory}


\author{Michael R. Douglas}
\email{douglas@max2.physics.sunysb.edu}
\altaffiliation{I.H.E.S.}

\author{Luca Mazzucato}
\email{mazzu@max2.physics.sunysb.edu}
%\thanks{}

\author{Shlomo S. Razamat}
\email{razamat@max2.physics.sunysb.edu}

\affiliation{Simons Center for Geometry and Physics and YITP\\
Stony Brook University\\
Stony Brook, NY 11794 USA\\}

\date{\today}

\begin{abstract}
We derive a holographic dual description of free quantum field theory in arbitrary
dimensions, by reinterpreting the exact renormalization group, to obtain a
higher spin gravity theory of the general type which had been proposed and studied
as a dual theory.  We show that the dual theory reproduces all correlation functions. 
\end{abstract}

%\keywords{}

\maketitle

\section{Introduction}

One of the most striking and unexpected discoveries of the 1994-98 ``second
superstring revolution'' was the AdS/CFT correspondence 
\cite{Maldacena:1997re},
according to which ${\mathcal N}=4$ supersymmetric Yang-Mills theory in four dimensions
is dual to type IIb superstring theory on $AdS_5\times S^5$.  Since then, the
correspondence has been much generalized  and has 
found many applications, especially in providing simple 
models exhibiting nonperturbative physical phenomena such as confinement, dissipation and
quantum phase transitions. 
However, despite a good deal of work, the microscopic workings of the duality
are not well understood.  In no case has there been a first principles derivation.

In this work, we derive a gravity dual to free field theory.
Free scalar field theory is conjectured~\cite{Sundborg:2000wp,Klebanov:2002ja} to 
be holographically dual to higher spin gravity
as developed by M.~Vasiliev and other authors~\cite{Bekaert:2005vh},
and nontrivial checks of this conjecture were made 
in~\cite{Giombi:2009wh,Koch:2010cy}.  By standard large $N$ arguments,
the same dual formulation should describe the large $N$ limit of the $O(N)$ model 
as well~\cite{Klebanov:2002ja}.
  
It is widely believed that AdS/CFT is at heart a geometric reformulation of the renormalization group (RG),
in which the renormalization scale becomes an extra `radial' dimension.  Various explanations of
this idea have been given, such as the holographic RG~\cite{de Boer:1999xf}, and a mixed 
holographic/Wilsonian RG ~\cite{Heemskerk:2010hk}, while attempts at a precise
reformulation were made in~\cite{Lee:2009ij}.   Here we begin by reviewing the exact RG.

\medskip 

\noindent {\bf {Exact RG equations:}} We study the theory of $N$ free complex scalar fields, 
denoted $\phi^A(x)$, in $D$  dimensions.
The bare action will be a sum of a standard (two derivative) kinetic term, and a
$U(N)$-invariant interaction
term with arbitrary position and momentum dependence,
\be\nonumber
S = \sum_A \int d^Dx\ |\partial\phi^A(x)|^2 - \int d^Dx d^Dy\ B(x,y)\bar\phi^A(x)\phi^A(y) .
\ee

Going to momentum variables $p,q$, and writing $\phi^A(p)$ for the Fourier transform,
the Wilsonian effective action at energy scale $\L$ is 
\be\label{action}
S=
\int d^D p\, d^D q\biggl\{P(p,q) - B(p,q)\biggr\} \bar\phi^A(q)\phi^A(p)\,,
\ee
with a cutoff kinetic term  
\be 
 P={p^2K^{-1}(p^2/\L^2)} \delta^{(D)}(p-q)\, .
%\frac{1}{p^2} \, K(p^2/\L^2)\,,
\ee
The cutoff function $K$ is chosen so
that the propagator vanishes for high momenta and goes to $1/p^2$ for small momenta.
We also define
\be\label{defalpha}
\a= \frac{d_\L K(p^2/\L^2)\,}{p^2}\;\delta^{(D)}(p-q)\,.
\ee 

Applying the standard derivation of Wick's theorem from the functional integral, and taking
a derivative with respect to $\L$, one obtains an exact RG equation~\cite{Polchinski:1983gv}.
Since the theory is free, under RG flow the effective action remains quadratic in the fields. 
The coupling $B$ flows as
\bea\label{rgeq}
&&d_\L B(p,q)=\\
&&\qquad\qquad- \int d^Ds \frac{1}{s^2}\frac{\d K(s^2/\L^2)}{\d\L} \, B(p,s)\,B(s,q)\, .\nonumber 
\eea
There is also a constant term $F$,
satisfying the flow equation
\be\label{Fder}
d_\L F= N\,\int d^Dp d^Dq\, \a(p,q) \left(P(q,p)+B(q,p)\right) .
\ee 
Integrating this flow down to $\Lambda\rightarrow 0$, one obtains the free energy.

Connected correlation functions of the bilocal operator $\bar\phi^A(x)\phi^A(y)$ can be obtained
as functional derivatives of the free energy with respect to $B(x,y)$.  Of course, since this is a
free theory, there is an explicit expansion for it,
\bea \label{neumann-exp}
F &=& - N\; \Tr \log (P-B) \nonumber\\
 &=& - N\; \Tr \log P + N \sum_{n\ge 1} \frac{1}{n} \Tr \left(P^{-1}\cdot B\right)^n ,
\eea
where $P^{-1}$ is the Green function, and $\Tr$ and $\cdot$ represent integration and products
of kernels.  Terms in this expansion correspond to one-loop diagrams with vertices taken from
the interaction $B$.  Our question is, what does this have to do with anti-de Sitter space and gravity?

\section{RG as equations of motion on AdS}

We now rewrite the RG flow equation~\eqref{rgeq} as an equation of motion for fields propagating
in an $AdS_{D+1}$ space with radial coordinate
\be \label{def-r}
r = \frac{1}{\L} .
\ee
The other coordinates of $AdS_{D+1}$
are `reference coordinates' whose relation to the original space-time coordinates
will be explained below.
The fields on $AdS_{D+1}$ will be a field $B$ derived from the couplings $B$, and
a connection $W$ in the `higher spin gauge algebra $hs(D-1,2)$, also to be defined below.

The first step is to reformulate~\eqref{rgeq} in terms of operators $B$ and $\alpha$ 
with a simple multiplication law.
To make contact with higher spin gravity as presented in~\cite{Bekaert:2005vh},
we will use an explicit representation in 
which operators are represented by symbols and operator products are represented by the
Moyal star product.  From now on we will discuss dimensionless $B$, i.e.  $B\to B(\frac{p}{\L},\frac{q}{\L})\L^{2-D}$.
We Taylor
expand the sources $B(p/\L,q/\L)$ in the momentum variables
\bea %\label{B-exp}
&&B(p/\L,q/\L)=\sum_{s,t=0}^\infty \,\L^{-s-t}\,B_{a_1\ldots a_s,b_1\ldots b_t}p^{a_1}\!\!\!\ldots  p^{a_s}q^{b_1}\!\!\!\ldots  q^{b_t}\nonumber\\
&&\qquad\qquad \equiv \L^{-s-t}\,B_{\underline{st}}\,p^{\underline{s}}\,q^{\underline{t}}\ ,\nonumber
\eea
where the indices $a_i$ and $b_i$ take values in $\{0,\dots, D-1\}$.  We then define
\be \label{a-exp}
\a^{\underline{st}} = \L^{2-D-s-t}\,\int d^Dp \int d^Dq\ \a(p,q) p^{\underline{s}}\,q^{\underline{t}}\,,
\ee
so that the RG flow equation~\eqref{rgeq} becomes
\bea\label{radialeom}
&&\frac{d}{d\L} B_{\underline{st}}=-B_{\underline{si}}\,\a^{\underline{ij}}\,B_{\underline{jt}}+\L^{-1}(s+t+2d_\phi)\,B_{\underline{st}}\,,
\eea where $d_\phi=\frac{D-2}{2}$ is the conformal dimension of $\phi^A$. 

Now, an RG flow equation expresses an identification between theories with the 
same physics, written in terms of actions defined at infinitesimally different energy scales.
Mathematically, such an infinitesimal relation should be expressed by a connection
on the space of actions.  In fact it is simple to reinterpret \eqref{radialeom} in this way.
Define a connection one-form, whose only component is
\bea \label{postulate-zero}
(W_\Lambda)_{\underline{s}}{}^{\underline{j}} &=& B_{\underline{si}}\,\a^{\underline{ij}} - s\L^{-1}  \delta_{\underline{s}}{}^{\underline{j}} \ ,
\\ \nonumber
(\widetilde W_\Lambda)^{\underline{k}}{}_{\underline{t}} &=&
  -t\L^{-1}  \delta^{\underline{k}}{}_{\underline{t}} \ ,
\eea
then \eqref{radialeom} becomes
\be
0 = \frac{d}{d\L} \Lambda^{-2d_\phi} B_{\underline{st}} + (W_\Lambda)_{\underline{s}}{}^{\underline{j}} \,\Lambda^{-2d_\phi}B_{\underline{jt}}
 + \Lambda^{-2d_\phi}B_{\underline{sj}}  (\widetilde W_\Lambda)^{\underline{j}}{}_{\underline{t}}\, .
\ee
Our description of the action also depends on a
choice of spatial reference point.  By rewriting the position space interaction
term as
\be \nonumber
\int d^Dx' d^Dx\ B(a+x,a+x')\bar\phi^A(a+x) \phi^A(a+x') ,
\ee
one sees that an overall shift symmetry $x\to x+a$ acts as
\be\label{spatialeom}
\frac{d}{da_i}\,B_{\underline{st}}p^{\underline{s}}q^{\underline{t}}=i
(p^i-q^i)\, B_{\underline{st}}p^{\underline{s}}q^{\underline{t}}. 
\ee
We will also interpret this as a connection on the space of actions.  Thus,
we need to reinterpret 
the right hand sides of \eqref{radialeom} and~\eqref{spatialeom} as the action of
a gauge algebra.  Mathematically, this will be an algebra of pseudodifferential operators.
But here, motivated by the eventual contact with higher spin gravity,
we will define the gauge algebra as the Lie algebra associated to an associative algebra,
defined by a Moyal star product.

We introduce oscillators (formal auxiliary variables) $y^\a,\,\bar y_\a,\,z^\a$
and $\bar z_\a$, where $\a\in\{\bullet,r,0,1,\dots, D-1\}$, satisfying the Moyal star product \cite{Bekaert:2005vh}
\bea\label{aux}
&&(f*g)(z,y)=\\
&&{1\over\pi^{2(D+2)}}\int dsdte^{-2s\cdot\bar t-2\bar s\cdot t}f(z+s,y+s)g(z-t,y+t) \ \nonumber.
\eea
%\be \label{aux}
%[\bar y_\a,\,y^\b]_*=\eta^\b_\a,\quad [\bar z_\a,\,z^\b]_*=-\eta^\b_\a\,,
%\ee 
The metric on this auxiliary space is ${\hat\eta}_\alpha^\beta=(-1,1,\eta)$, where $\eta$ is the metric on the original flat space-time. We further define for $a\in\{0,..,D-1\}$
 \bea
&&y^a=Y^a+ Z^a,\qquad 
\bar y_a = \half(\bar Y_a- \bar Z_a),\\
&&z^a=Z^a- Y^a,\qquad 
\bar z_a = \half(\bar Y_a+ \bar Z_a)\,.\nonumber
\eea
Then the field $B$ and the kernel $\a$ are defined to be functions of the auxiliary variables,
derived from the coupling $B$ and cutoff propagator variation $\alpha$ as (we trade momentum $p$ with $iY/r$ and $q$ with $-iZ/r$)
\bea\label{defBaux}
&&B(y,z,\bar y,\bar z)= i^{s-t}r^{D-2}\,B_{\underline{st}}\,Y^{\underline{s}} \,Z^{\underline{t}}\, e^{-Y\bar Y-Z\bar Z}\,(\bar z_{r}-\bar z_{\bullet})^{s+t}\,,\nonumber\\
&&\a_\mu(y,z,\bar y,\bar z)=\\
&&\qquad-\frac{(-i)^{t-s}}{s!t!} r^{-D}\,\a_\mu^{\underline{st}}\,\bar Y_{\underline{t}} \,\bar Z_{\underline{s}}\, e^{-Y\bar Y-Z\bar Z}\,(\bar z_{r}-\bar z_{\bullet})^{-s-t}\, ,\nonumber
\eea
where $\mu\in\{r,0,\ldots,D-1\}$. To rewrite the equations we just derived, from a starting point with an explicit translation action
on the coordinates, we can take $\a_{r}=\a$ and $\a_a=0$.  This choice can be generalized as
will emerge below.

\medskip 

\noindent {\bf {Standard connection on AdS:}} 
The group of linear transformations on each set of auxiliary variables preserving
the metric $\eta^\b_\a$ is $SO(D-1,2)$, the group of isometries of $AdS_{D+1}$.
As is well known, we can
represent the corresponding Lie algebra as star commutators
with generators which are quadratic functions of the oscillators.  
The dilatation and the translation generators are 
\bea\label{P-algebra}
P_{r}&=&\bar z_{r} z^{\bullet}-\bar z_{\bullet}z^{r},\\
P_a&=&\bar z_a\,(z^{\bullet}-z^{r})-(\bar z_{\bullet}-\bar z_{r})\,z^a\,.\nonumber
\eea
These can be used to define a connection
\be\label{connection}
W^{(0)}_\mu = \frac{1}{r}\, P_\mu\,, 
\ee
which due to the commutation relations satisfied by $P_\mu$ is flat
\be 
dW^{(0)}+W^{(0)}\wedge*W^{(0)}=0\,.
\ee 

Now, there is a well-known way to rewrite theories of gravity, not in terms of a 
metric, but in terms of a connection acting on the frame bundle (see for example \cite{Witten:1988hc}).
The connection \eqref{connection} is the one corresponding
to the $AdS_{D+1}$ metric in the Poincare patch,
\be
ds^2=\frac{dr^2+dx^adx^a}{r^2} \,.
\ee 
The standard formulations of higher spin gravity are also in terms of a connection, now living in
an infinite dimensional algebra $hs(D-1,2)$ which contains $SO(D-1,2)$.  By using
\eqref{P-algebra} to define \eqref{connection}, we have postulated just this structure.  Of course
this is just kinematic, a particular way to describe  $AdS_{D+1}$.

\medskip 

\noindent {\bf {RG as connection on AdS:}} Returning to~\eqref{defBaux},  
the field $B$ was dressed with the auxiliary variables in a way so that $[W^{(0)},B]_*$ reproduces
the linear term in $B$ in the RG equations~\eqref{radialeom} and~\eqref{spatialeom}. In particular
\bea
&&[W^{(0)}_r, (\bar z_r-\bar z_\bullet)^{s+t}]_*=\frac{s+t}{r}(\bar z_r-\bar z_\bullet)^{s+t},\;\\
&& [W^{(0)}_a, B]_*=\frac{y^a}{r} (\bar z_r-\bar z_{\bullet}) B
 - \frac{z_r-z_{\bullet}}{r} \frac{\partial}{\partial z^a} B.\nonumber
\eea
The last term in the second of these equations does not appear in ~\eqref{spatialeom}, since in the field theory the components of $B$ do not have $\bullet$ and $r$ indices.
The cut-off kernel $\a$ we have introduced is consistent with momentum conservation. 
With the notations introduced
above, this fact translates into the equation
\be\label{alfFlat}
d\a+W^{(0)}\wedge*\,\a+\a\wedge*\,W^{(0)}=0\,. 
\ee 
In fact, this equation can be satisfied more generally, which allows using 
position dependent cutoffs, or working on space-times without translation invariance. 

Finally, we define a fluctuation of the connection
\be\label{defWB}
{(\delta\widetilde W_\mu)^{\underline{p}}}_{\underline{q}} =0,\qquad
{( \delta W_\mu)_{\underline{q}}}^{\underline{p}} =B_{\underline{qr}}\,{\a^{\underline{rp}}}_\mu\,.
\ee 
With these definitions, the RG equations take the appealing form 
\be\label{starB}
\frac{d}{dx_\mu}\,B+W_\mu*\, B- B*\, \widetilde W_\mu=0\,, 
\ee where $W=W^{(0)}+\delta W$ and $\widetilde W=W^{(0)}+\delta \widetilde W$
Moreover, by right star multiplication of the above equation 
by $\a_\nu$ and antisymmetrization with respect to the space-time indices,
we obtain that the total connection is flat 
\be\label{flatW}
dW+W\wedge *\, W=0\,,\quad d\widetilde W+\widetilde W\wedge *\, \widetilde W=0\,.
\ee  

The equations \eqref{starB} and \eqref{flatW} admit the standard gauge transformations
\bea
&&\delta W=d\e+\left[W,\,\e\right]_*\,,\qquad
\delta \widetilde W=d\widetilde\e-\left[\widetilde W,\,\widetilde\e\right]_*\,,\\
&&\delta B=B*\,\widetilde\e-\e*\, B\;\;\,. \nonumber
\eea
and this is a gauge theory with gauge algebra $hs(D-1,2)$ as defined above.  In gauge theory
terms, these equations express the covariant
constancy of a bulk field $B$ on $AdS_{D+1}$ under transport by a flat connection $W$. 
Conceptually, their field theoretic origin is clear.  The AdS space parameterizes choices which
must be made to define the RG; an infinitesimal relation between equivalent
actions should be expressed by transport by a connection.  
If we vary the RG scale and the reference point along a closed loop in AdS
we must recover the same action, so the connection must be flat.

In the standard formulations of higher spin
gravity \cite{Bekaert:2005vh}, one has the same connection $W_\mu$,
and the higher spin fields are obtained by expanding it in the auxiliary variables. 
The equation \eqref{flatW} is an equation of motion, whose
linearization describes propagation of higher spin fields in $AdS_{D+1}$.
The field $B$ encodes, among others, the matter field coupled to the higher spin gauge fields and satisfies \eqref{starB} or a similar
equation of motion.
Thus, we have reformulated the RG flow for $D$-dimensional free field theory in
the terms of higher spin gravity in $D+1$ dimensions.

Finally, we obtain the solutions which correspond to RG flows by imposing \eqref{defWB}, 
{\it i.e.} the relation $W=B*\a$.  This relation is not gauge invariant; while formally one
can postulate a transformation law for $\a$ which would make it so, this requires taking
the inverse $B^{-1}$, which may not exist.  As a relation between one-forms, some of it
may correspond to a gauge fixing condition, while other parts have suggestive analogs in
higher spin gravity.  

\medskip 

\noindent {\bf {Action}:} One may reasonably ask what action gives rise to these equations.  Actually this problem 
has not been solved for the standard higher spin gravity theories; the equations
\eqref{starB} and \eqref{flatW} do not naturally come from an action in dimensions $D+1\ge 4$.
One can still postulate an action whose variational equations include these equations,
most simply by postulating a Lagrange multiplier $\lambda$ for each equation.  Such an action will be
zero evaluated on a solution, {\it i.e.} on-shell.

From our derivation, 
the on-shell action of our dual theory is the sum of such a zero on-shell action, and the
holonomy of the $U(1)$ part of the connection,
\be\label{freecov}
S_{bulk} = -N\, \Tr\,\int dx^\mu\, \delta W_\mu\, + S_{on-shell},
\ee 
integrated along a contour which runs from a point on
 the boundary to $r=\infty$.  This follows from the expression
\eqref{Fder} for the free energy of the field theory and
the identification~\eqref{defWB}. 
Mathematically, it follows from the identification of $\Tr W_\mu$ as a connection on the
determinant line bundle of operators $-P+B$.

Since the connection is flat, one can deform the contour and obtain the same result.  
One can also vary the choice of base point
on the boundary; this corresponds to a gauge transformation.

\medskip 

\noindent {\bf {Correlators:}} Correlators in $AdS/CFT$ correspondence are obtained by varying the bulk action
\eqref{freecov} with respect to the sources.  We take the contour to extend between two
boundary points and take the limit of one point going to infinity.

When evaluated on-shell the action of our bulk theory is equal to a holonomy. 
The gauge transformations
change $W_\mu$ and thus change the sources. To compute the holonomy one has to evaluate the connection $W_\mu$ in the bulk.

This task can be achieved by solving the equations of motion perturbatively in the sources (i.e. by Witten diagrams).
\begin{figure}[htbp]
\begin{center}
\centerline{\includegraphics[scale=0.3]{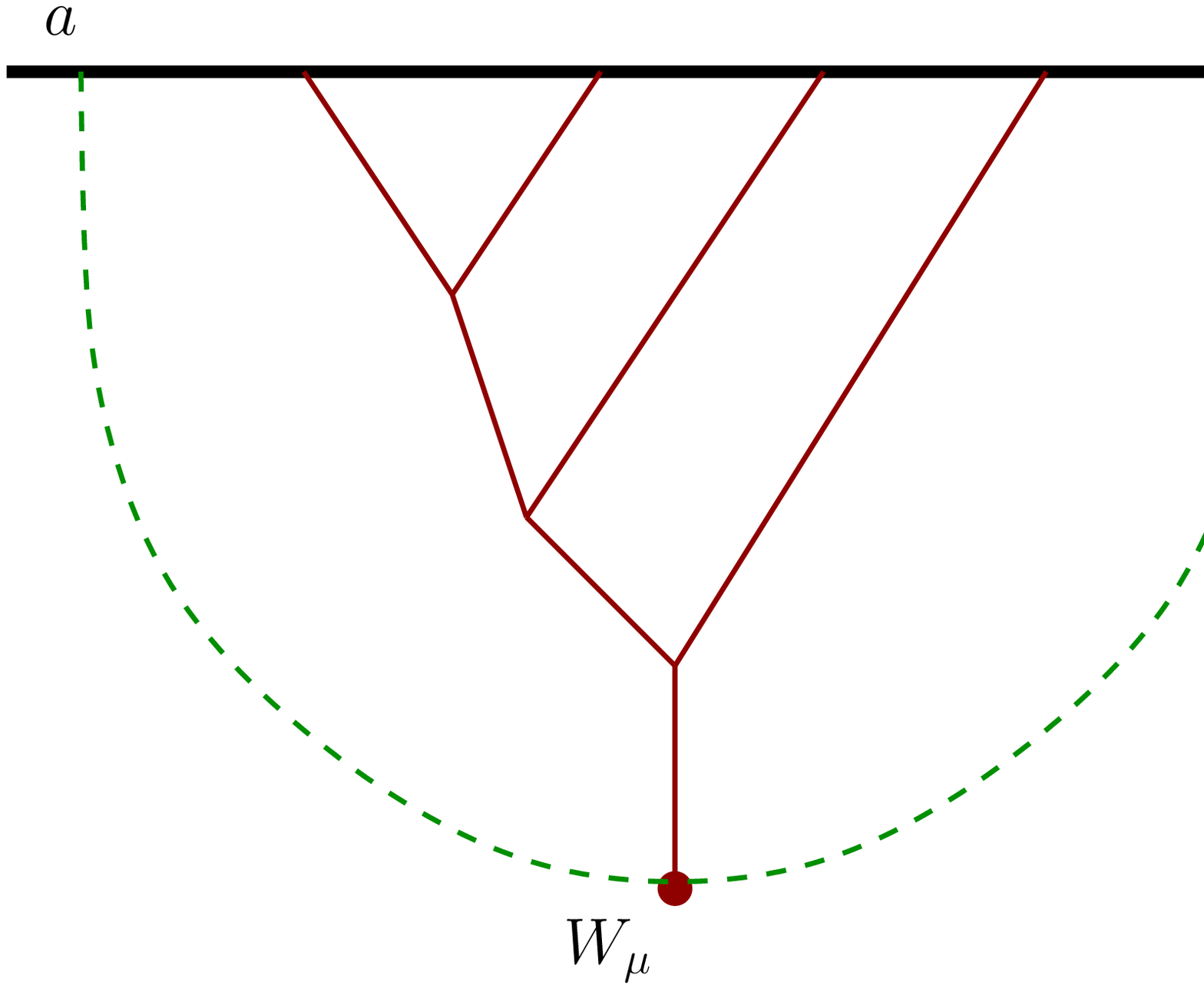}}
\end{center}
\begin{center}
\caption{A graphic representation of the calculation of a four point correlator.
The dashed line is the holonomy contour. $a$ and $b$ are two points on the boundary; one of which is taken to infinity. 
The (brown) lines represent boundary-to-bulk and bulk-to-bulk propagators.} 
\label{holon}
\end{center}
\end{figure} 
We define a perturbative expansion of $B$ as
\be\label{expanB}
B=B^{(0)}+ B^{(1)}+B^{(2)}+\dots\,.
\ee Here $B^{(0)}$ is the solution to the linearized equation~\eqref{starB},
\be\label{linB} 
\frac{d}{dx^\mu}B^{(0)}+W^{(0)}_\mu *B^{(0)}- B^{(0)}* \widetilde W^{(0)}_\mu=0\,.
\ee  One defines the boundary-to-bulk propagator, $K(x,x',r)$ as a solution to the above equation with $\delta$-function boundary condition, and the bulk-to-bulk
propagator, ${G^\nu}(x,x',r,r')$,  as a solution with $\delta^{(D)}(x-x')\delta(r-r'){\eta^\nu}_\mu$ source 
 on the right-hand side. 
 
The solution to the linearized equations of motion for the fluctuations (\ref{defWB}) is given by 
$\delta B=g^{-1}*b*dg$ with
\be\label{groupel}
g(x;z)=P\exp_*\left(-\int_x^{x_0}W^{(0)}_\mu {dx'}^\mu\right) \ ,
\ee
with $W^{(0)}$ as in (\ref{connection}). 
By taking a straight contour with base point $x_0^\mu=(r_0,x_0^a)$ we get rid of the path ordering
\be
g(x;y)=\exp_*\left(P_\mu\frac{(x-x_0)^\mu}{r-r_0}\ln{r\over r_0}\right) \ .
\ee

 The boundary conditions are given by specifying the boundary sources $\hat B$. Then,
\bea
&&B^{(0)}(x,r)=\int d^Dx'\, K(x,x',r)* \hat B(x'),\\
&&B^{(1)}(x,r)=\nonumber\\
&&-\int dr'd^Dx'\,{G^{\nu}}(x,x',r,r')* \left[B^{(0)}*\a_\nu*B^{(0)}\right](x',r'),\nonumber    
\eea 
and the higher corrections are obtained in a similar manner. 
Since $\a$ appears in the interaction vertex, it should be
chosen to be regular at $p=0$, {\it e.g.}  $\a\sim exp(-p^2/\L^2)/\L^3$.
The correlators are then given by varying the holonomy integral~\eqref{freecov} with respect to $\hat B$ and by construction reproduce
the field theory results. This procedure is illustrated in figure~\ref{holon}.  Summing the
diagrams and doing the $r$ integrals, one reproduces the expansion \eqref{neumann-exp},
thus answering the question of our introduction.

\section{Discussion}

Starting from free bosonic field theory, with an arbitrary position
or momentum-dependent kinetic term (dispersion relation), we have derived a dual
description as a higher spin gravity in anti-de Sitter space, and argued that it can reproduce all
correlation functions.  Although the higher spin gravity we arrived at is not of the standard form,
it contains the structure tested in explicit comparisons such as \cite{Giombi:2009wh} and seems
as well motivated from this point of view as the standard theories.  One point which could be
improved is to rephrase the relation \eqref{defWB} in a more covariant way.

One of the outstanding conceptual questions about AdS/CFT is to understand the relation between
the two dual space-times.  There is a common though not universal belief that the relation is
nonlocal away from the boundary and that any microscopic derivation must include some sort of nonlocal transformation.  On the other hand, our derivation did not do this; rather, we moved all of the nonlocality
into the interactions and dependence on the auxiliary variables.  If the same can be done in gauge theory,
in which higher derivative operators have large anomalous dimension at strong coupling, then it 
seems reasonable to look for a local understanding of the duality there as well. 

In the end, the significance of this result depends on the extent to which these ideas and techniques apply to
interacting theories.  The only case for which this generalization will be direct is the
interacting $O(N)$ model, as we will discuss elsewhere. 

\medskip
\noindent{\bf Acknowledgments}:~
We would like to thank L.~Rastelli, D.~Sorokin, B.~van~Rees, S.~J.~Rey, and E.~Witten   for very useful discussions.
We also are grateful to O.~J.~Rosten for comments on the first version of the paper.
This research was supported in part by DOE grant DE-FG02-92ER40697 and by NSF grant PHY-0653351-001. Any
opinions, findings, and  conclusions or recommendations expressed in this
material are those of the authors and do not necessarily reflect the views of the National
Science Foundation.



\begin{thebibliography}{99}

%\cite{Maldacena:1997re}
\bibitem{Maldacena:1997re}
  J.~M.~Maldacena,
  %``The large N limit of superconformal field theories and supergravity,''
  Adv.\ Theor.\ Math.\ Phys.\  {\bf 2}, 231 (1998)
  [Int.\ J.\ Theor.\ Phys.\  {\bf 38}, 1113 (1999)]
  [arXiv:hep-th/9711200];
  %%CITATION = IJTPB,38,1113;%%
  S.~S.~Gubser, I.~R.~Klebanov and A.~M.~Polyakov,
  %``Gauge theory correlators from non-critical string theory,''
  Phys.\ Lett.\  B {\bf 428}, 105 (1998)
  [arXiv:hep-th/9802109];
  %%CITATION = PHLTA,B428,105;%%
  E.~Witten,
  %``Anti-de Sitter space and holography,''
  Adv.\ Theor.\ Math.\ Phys.\  {\bf 2}, 253 (1998)
  [arXiv:hep-th/9802150].
  %%CITATION = 00203,2,253;%%

%\cite{Sundborg:2000wp}
\bibitem{Sundborg:2000wp}
  B.~Sundborg,
  %``Stringy gravity, interacting tensionless strings and massless higher
  %spins,''
  Nucl.\ Phys.\ Proc.\ Suppl.\  {\bf 102}, 113 (2001)
  [arXiv:hep-th/0103247];
  %%CITATION = NUPHZ,102,113;%%
  E.~Sezgin and P.~Sundell,
  %``Massless higher spins and holography,''
  Nucl.\ Phys.\  B {\bf 644}, 303 (2002)
  [Erratum-ibid.\  B {\bf 660}, 403 (2003)]
  [arXiv:hep-th/0205131].
  %%CITATION = NUPHA,B644,303;%%
  
%\cite{Klebanov:2002ja}
\bibitem{Klebanov:2002ja}
  I.~R.~Klebanov and A.~M.~Polyakov,
  %``AdS dual of the critical O(N) vector model,''
  Phys.\ Lett.\  B {\bf 550}, 213 (2002)
  [arXiv:hep-th/0210114].
  %%CITATION = PHLTA,B550,213;%%

 %\cite{Bekaert:2005vh}
\bibitem{Bekaert:2005vh}
  M.~A.~Vasiliev,
  %``Higher spin gauge theories: Star-product and AdS space,''
  arXiv:hep-th/9910096;
  %%CITATION = HEP-TH/9910096;%%
  X.~Bekaert, S.~Cnockaert, C.~Iazeolla and M.~A.~Vasiliev,
  %``Nonlinear higher spin theories in various dimensions,''
  arXiv:hep-th/0503128.
  %%CITATION = HEP-TH/0503128;%%

 %\cite{Giombi:2009wh}
\bibitem{Giombi:2009wh}
  S.~Giombi and X.~Yin,
  %``Higher Spin Gauge Theory and Holography: The Three-Point Functions,''
  JHEP {\bf 1009}, 115 (2010)
  [arXiv:0912.3462 [hep-th]];
  %``Higher Spins in AdS and Twistorial Holography,''
  arXiv:1004.3736 [hep-th].
  %%CITATION = ARXIV:1004.3736;%%

%\cite{Koch:2010cy}
\bibitem{Koch:2010cy}
  R.~d.~M.~Koch, A.~Jevicki, K.~Jin and J.~P.~Rodrigues,
  %``AdS_4/CFT_3 Construction from Collective Fields,''
  arXiv:1008.0633 [hep-th];
  M.~Henneaux and S.~J.~Rey,
  %``Nonlinear W(infinity) Algebra as Asymptotic Symmetry of Three-Dimensional
  %Higher Spin Anti-de Sitter Gravity,''
  arXiv:1008.4579 [hep-th];
  %%CITATION = ARXIV:1008.4579;%%
  A.~Campoleoni, S.~Fredenhagen, S.~Pfenninger and S.~Theisen,
  %``Asymptotic symmetries of three-dimensional gravity coupled to higher-spin
  %fields,''
  JHEP {\bf 1011}, 007 (2010)
  [arXiv:1008.4744 [hep-th]];
  %%CITATION = JHEPA,1011,007;%%
  %%CITATION = ARXIV:1008.0633;%%
  M.~R.~Gaberdiel and R.~Gopakumar,
  %``An AdS_3 Dual for Minimal Model CFTs,''
  arXiv:1011.2986 [hep-th].
  %%CITATION = ARXIV:1011.2986;%%
  
  
%\cite{de Boer:1999xf}
\bibitem{de Boer:1999xf}
E.~T.~Akhmedov,
  %``A remark on the AdS/CFT correspondence and the renormalization group
  %flow,''
  Phys.\ Lett.\  B {\bf 442}, 152 (1998)
  [arXiv:hep-th/9806217];
  J.~de Boer, E.~P.~Verlinde and H.~L.~Verlinde,
  %``On the holographic renormalization group,''
  JHEP {\bf 0008}, 003 (2000)
  [arXiv:hep-th/9912012].
  %%CITATION = JHEPA,0008,003;%%
  

  
%\cite{Heemskerk:2010hk}
\bibitem{Heemskerk:2010hk}
  I.~Heemskerk and J.~Polchinski,
  %``Holographic and Wilsonian Renormalization Groups,''
  arXiv:1010.1264 [hep-th];
  %%CITATION = ARXIV:1010.1264;%%
T.~Faulkner, H.~Liu and M.~Rangamani,
  %``Integrating out geometry: Holographic Wilsonian RG and the membrane
  %paradigm,''
  arXiv:1010.4036 [hep-th].
  %%CITATION = ARXIV:1010.4036;%%
  
  
%\cite{Lee:2009ij}
\bibitem{Lee:2009ij}
  E.~T.~Akhmedov and E.~T.~Musaev,
  %``An exact result for Wilsonian and Holographic renormalization group,''
  Phys.\ Rev.\  D {\bf 81}, 085010 (2010)
  [arXiv:1001.4067 [hep-th]];
  S.~S.~Lee,
  %``Holographic description of quantum field theory,''
  Nucl.\ Phys.\  B {\bf 832}, 567 (2010)
  [arXiv:0912.5223 [hep-th]]; arXiv:1011.1474 [hep-th].
  %%CITATION = NUPHA,B832,567;%%


  
  
  %\cite{Polchinski:1983gv}
\bibitem{Polchinski:1983gv}
  J.~Polchinski,
  %``Renormalization And Effective Lagrangians,''
  Nucl.\ Phys.\  B {\bf 231}, 269 (1984).
  %%CITATION = NUPHA,B231,269;%%
  O.~J.~Rosten,
  %``Fundamentals of the Exact Renormalization Group,''
  arXiv:1003.1366 [hep-th].
  %%CITATION = ARXIV:1003.1366;%%
  
    %\cite{Witten:1988hc}
\bibitem{Witten:1988hc}
  E.~Witten,
  %``(2+1)-Dimensional Gravity as an Exactly Soluble System,''
  Nucl.\ Phys.\  B {\bf 311}, 46 (1988).



\end{thebibliography}
\end{document}